\documentclass[aps,pra,reprint]{revtex4-2}
\usepackage{graphicx, amsmath, amssymb, enumitem, hyperref}
\usepackage[caption=false]{subfig}

\begin{document}


\newcommand{\comb}[2]{{\begin{pmatrix} #1 \\ #2 \end{pmatrix}}}
\newcommand{\braket}[2]{{\left\langle #1 \middle| #2 \right\rangle}}
\newcommand{\bra}[1]{{\left\langle #1 \right|}}
\newcommand{\ket}[1]{{\left| #1 \right\rangle}}
\newcommand{\ketbra}[2]{{\left| #1 \middle\rangle \middle \langle #2 \right|}}

\newcommand{\fref}[1]{Fig.~\ref{#1}}
\newcommand{\Fref}[1]{Figure~\ref{#1}}
\newcommand{\sref}[1]{Sec.~\ref{#1}}
\newcommand{\tref}[1]{Table~\ref{#1}}


\title{Quantum Search with the Signless Laplacian}

\author{Molly E. McLaughlin}
	\email{mollymclaughlin@creighton.edu}
	\affiliation{Department of Physics, Creighton University, 2500 California Plaza, Omaha, NE 68178}
\author{Thomas G. Wong}
	\email{thomaswong@creighton.edu}
	\affiliation{Department of Physics, Creighton University, 2500 California Plaza, Omaha, NE 68178}

\begin{abstract}
	Continuous-time quantum walks are typically effected by either the discrete Laplacian or the adjacency matrix. In this paper, we explore a third option: the signless Laplacian, which has applications in algebraic graph theory and may arise in layered antiferromagnetic materials. We explore spatial search on the complete bipartite graph, which is generally irregular and breaks the equivalence of the three quantum walks for regular graphs, and where the search oracle breaks the equivalence of the Laplacian and signless Laplacian quantum walks on bipartite graphs without the oracle. We prove that a uniform superposition over all the vertices of the graph partially evolves to the marked vertices in one partite set, with the choice of set depending on the jumping rate of the quantum walk. We boost this success probability to 1 by proving that a particular non-uniform initial state completely evolves to the marked vertices in one partite set, again depending on the jumping rate. For some parameter regimes, the signless Laplacian yields the fastest search algorithm of the three, suggesting that it could be a new tool for developing faster quantum algorithms.
\end{abstract}

\maketitle


\section{Introduction}

A continuous-time quantum walk is a quantum analogue of a continuous-time random walk, and it is a universal model for quantum computing \cite{Childs2009,Herrman2019}. In such a quantum walk, the state $\ket{\psi}$ is a superposition (or linear combination) over the $N$ vertices of a graph, i.e.,
\[ \ket{\psi(t)} = c_1(t) \ket{1} + c_2(t) \ket{2} + \dots + c_N(t) \ket{N}, \]
where $|c_i|^2$ is the probability of measuring the walker at vertex $i$, and $\ket{i}$ is the basis vector for vertex $i$. If the walker is found at vertex $i$, then the system is now in the state $\ket{i}$, i.e., the state ``collapsed.'' The state $\ket{\psi}$ evolves by Schr\"odinger's equation, the fundamental equation of quantum mechanics,
\begin{equation}
    \label{eq:Schrodinger}
    i \frac{d\ket{\psi}}{dt} = H \ket{\psi},
\end{equation}
where we have set the reduced Planck's constant $\hbar = 1$, and $H$ is the Hamiltonian, which is a Hermitian (self-adjoint) matrix that characterizes the total energy of the system and which respects the structure of the graph.

The Hamiltonian $H$ can take a variety of forms, depending on the system. For a quantum particle moving in continuous space by virtue of its kinetic energy, the Hamiltonian is proportional to the negative of Laplace's operator, i.e., $H = -\gamma \nabla^2$, where $\nabla^2 = \partial^2/\partial x^2 + \partial^2/\partial y^2 + \partial^2/\partial z^2$. In discrete space, Laplace's operator is replaced by the discrete Laplacian $L$, defined by
\begin{equation}
    \label{eq:Laplacian}
    L = A - D
\end{equation}
where $A$ is the adjacency matrix with
\[ A_{ij} = \begin{cases}
    1, & \text{if vertices $i$ and $j$ are adjacent}, \\
    0, & \text{otherwise},
\end{cases} \]
and $D$ is the degree matrix, which is a diagonal matrix where $D_{ii}$ equals the degree, or number of neighbors, of vertex $i$. The result of changing from continuous to discrete space by replacing $\nabla^2 \to L$ yields a \emph{Laplacian quantum walk} with Hamiltonian $H = -\gamma L$, where $\gamma$ is the jumping rate of the walk. Laplacian quantum walks are the basis for quantum algorithms for traversing decision trees \cite{FG1998a}, state transfer \cite{Alvir2016}, and spatial search \cite{CG2004}.

Another common kind of quantum walk is the \emph{adjacency quantum walk}, where the adjacency matrix appears without the degree matrix, so its Hamiltonian is $H = -\gamma A$. As we will review in \sref{section:spin-network}, whereas the Laplacian quantum walk arises in the Heisenberg (XYZ) model of a spin network in a single-excitation subspace, the adjacency quantum walk arises in the XY model. Adjacency quantum walks underpin quantum algorithms for traversing graphs \cite{Godsil2012}---including with an exponential speedup \cite{Childs2003}---spatial search \cite{Novo2015}, and solving boolean formulas \cite{FGG2008}.

\begin{figure}
\begin{center}
    \includegraphics{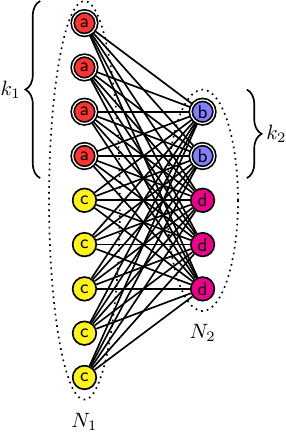}
    \caption{\label{fig:bipartite}A complete bipartite graph with $N_1 = 9$ and $N_2 = 5$ vertices in each partite set, of which $k_1 = 4$ and $k_2 = 2$ vertices are marked in the respective sets, as indicated by double circles. Identically evolving vertices are identically colored and labeled.}
\end{center}
\end{figure}

In this paper, we consider a third, lesser-studied type of quantum walk, one effected by the signless Laplacian
\begin{equation}
    \label{eq:signless-Laplacian}
    Q = A + D,
\end{equation}
where the Hamiltonian of the \emph{signless Laplacian quantum walk} is $H = -\gamma Q$. The signless Laplacian is used in spectral graph theory, such as to count the number of bipartite components \cite{Chung1997}. We will show in the next section that the signless Laplacian quantum walk can arise in the Heisenberg (XYZ) model for a spin network with certain couplings between spins, such in layer antiferromagnetic materials. While the signless Laplacian quantum walk has been explored for state transfer \cite{Christandl2004,Coutinho2015,Alvir2016,Tian2021,Zhang2022}, here we conduct the first \emph{algorithmic} investigation of the signless Laplacian quantum walk. Namely, we explore quantum search \cite{AA2005} on the complete bipartite graph with multiple marked vertices, an example of which is shown in \fref{fig:bipartite}. The partite sets have $N_1$ and $N_2$ vertices, of which $k_1$ and $k_2$ are ``marked'' in the respective sets. The total number of vertices is $N = N_1 + N_2$. The goal is to find a marked vertex by querying an oracle as few times as possible. In a continuous-time quantum walk, the oracle is an additional term the Hamiltonian, and so the task is to evolve the system for as little time as possible. Note the algorithmic component of this work is crucial---without the oracle, Laplacian and signless Laplacian quantum walks on bipartite graphs are identical in their probability distributions \cite{Alvir2016}, but as we will show, the walks differ within the context of searching, i.e., when an oracle is present. The complete bipartite graph was chosen because, due to the symmetry of the graph and the search algorithm, it is perhaps the easiest graph to analyze for which the three quantum walks differ.

\begin{table*}
\caption{\label{table:summary}Summary of search on the complete bipartite graph by Laplacian, adjacency, and signless Laplacian quantum walks with $N_1$ and $N_2$ large, and $N_1 \not\approx N_2$. When $N_1 \approx N_2$, all the walks behave like the adjacency quantum walk. The initial states $\ket{s} = \ket{s_L}$, $\ket{s_A}$, and $\ket{s_Q}$ are given in \tref{table:initial}. In the third, fifth, and sixth rows, $t_*$ refers to the runtime for each respective case.}
\begin{ruledtabular}
\begin{tabular}{ccccc}
	Quantum Walk & Jumping Rate & Evolution & Runtime & Success Probability \\[0.05in]
	\hline & \\[-0.1in]
    Laplacian & $\frac{1}{N_2}$ & $\ket{s} \to \ket{a}$ & $\frac{\pi}{2} \sqrt{\frac{N}{k_1}}$ & 1 \\[0.2in]
    Laplacian & $\frac{1}{N_1}$ & $\ket{s} \to \ket{b}$ & $\frac{\pi}{2} \sqrt{\frac{N}{k_2}}$ & 1 \\[0.1in]
	\hline & \\[-0.1in]
    Adjacency & $\frac{1}{\sqrt{N_1N_2}}$ & $\ket{s} \to \begin{array}{l} \frac{\sqrt{N_1}+\sqrt{N_2}}{\sqrt{2(N_1 + N_2)}} \left( \sqrt{\frac{k_1 N_2}{k_2 N_1 + k_1 N_2}} \ket{a} + \sqrt{\frac{k_2 N_1}{k_2 N_1 + k_1 N_2}} \ket{b} \right) \\ \quad + e^{it_*} \frac{-\sqrt{N_1}+\sqrt{N_2}}{2\sqrt{(N_1 + N_2)}} \left( -\ket{c} + \ket{d} \right) \end{array}$ & $\frac{\pi}{\sqrt{2}} \sqrt{\frac{N_1 N_2}{k_2 N_1 + k_1 N_2}}$ & $\frac{1}{2} + \frac{\sqrt{N_1N_2}}{N}$ \\[0.3in]
    Adjacency & $\frac{1}{\sqrt{N_1N_2}}$ & $\ket{s_A} \to \sqrt{\frac{k_1 N_2}{k_2 N_1 + k_1 N_2}} \ket{a} + \sqrt{\frac{k_2 N_1}{k_2 N_1 + k_1 N_2}} \ket{b}$ & $\frac{\pi}{\sqrt{2}} \sqrt{\frac{N_1 N_2}{k_2 N_1 + k_1 N_2}}$ & 1 \\[0.15in]
	\hline & \\[-0.1in]
    Signless Laplacian & $\frac{1}{N_1}$ & $\ket{s} \to \begin{array}{l} 2i \frac{\sqrt{N_1N_2}}{N} e^{i(1+N_2/N_1)t_*} \ket{a} + \frac{\sqrt{N_1}(N_1 - N_2)}{N^{3/2}} \ket{c} \\ \quad - \frac{\sqrt{N_2}(N_1 - N_2)}{N^{3/2}} \ket{d} \end{array}$ & $\frac{\pi}{2} \sqrt{\frac{N_1 N}{k_1N_2}}$ & $\frac{4N_1N_2}{N^2}$ \\[0.2in]
    Signless Laplacian & $\frac{1}{N_2}$ & $\ket{s} \to \begin{array}{l} 2i \frac{\sqrt{N_1N_2}}{N} e^{i(1+N_1/N_2)t_*} \ket{b} - \frac{\sqrt{N_1}(N_2 - N_1)}{N^{3/2}} \ket{c} \\ \quad + \frac{\sqrt{N_2}(N_2 - N_1)}{N^{3/2}} \ket{d} \end{array}$ & $\frac{\pi}{2} \sqrt{\frac{N_2 N}{k_2N_1}}$ & $\frac{4N_1N_2}{N^2}$ \\[0.2in]
    Signless Laplacian & $\frac{1}{N_1}$ & $\ket{s_Q} \to \ket{a}$ & $\frac{\pi}{2} \sqrt{\frac{N_1 N}{k_1N_2}}$ & 1 \\[0.2in]
    Signless Laplacian & $\frac{1}{N_2}$ & $\ket{s_Q} \to \ket{b}$ & $\frac{\pi}{2} \sqrt{\frac{N_2 N}{k_2N_1}}$ & 1 \\[0.1in]
\end{tabular}
\end{ruledtabular}
\end{table*}

\begin{table*}
\caption{\label{table:initial}Initial states in continuous-time quantum walk search algorithms on the complete bipartite graph, which are eigenvectors of either the Laplacian, adjacency matrix, or signless Laplacian with some eigenvalue.}
\begin{ruledtabular}
\begin{tabular}{cccc}
	Initial State & Description & Eigenvector of... & Eigenvalue \\[0.05in]
	\hline & \\[-0.15in]
    $\displaystyle \ket{s} = \ket{s_L} = \frac{1}{\sqrt{N}} \sum_{i = 1}^N \ket{i}$ & $\begin{array}{c} \text{Uniform superposition over all vertices,} \\ \text{each vertex has amplitude }1/\sqrt{N}. \end{array}$ & $L$ & 0 \\[0.2in]
    $\displaystyle \ket{s_A} = \frac{1}{\sqrt{2N_1}} \sum_{i \in V_1} \ket{i} + \frac{1}{\sqrt{2N_2}} \sum_{i \in V_2} \ket{i}$ & $\begin{array}{c} \text{Each left vertex has amplitude }1/\sqrt{2N_1}, \\ \text{each right vertex has amplitude }1/\sqrt{2N_2}. \end{array}$ & $A$ & $\sqrt{N_1N_2}$ \\[0.2in]
    $\displaystyle \ket{s_Q} = \frac{1}{\sqrt{N}} \left( \sqrt{\frac{N_2}{N_1}} \sum_{i \in V_1} \ket{i} + \sqrt{\frac{N_1}{N_2}} \sum_{i \in V_2} \ket{i} \right)$ & $\begin{array}{c} \text{Each left vertex has amplitude }\sqrt{N_2/(N_1N)}, \\ \text{each right vertex has amplitude }\sqrt{N_1/(N_2N)}. \end{array}$& $Q$ & $N$ \\[0.1in]
\end{tabular}
\end{ruledtabular}
\end{table*}

Searching the complete bipartite graph with multiple marked vertices was explored for the Laplacian and adjacency quantum walks in \cite{Wong19}, and our investigation shows that the signless Laplacian quantum walk searches differently from the other two. \tref{table:summary} summarizes the results that we will be deriving. In the first two rows, with the Laplacian quantum walk, the jumping rate $\gamma$ could be chosen so that the system entirely evolves from a uniform superposition over all the vertices $\ket{s}$ to only the marked vertices in one partite set, while another value of $\gamma$ causes it to entirely evolve to only the marked vertices in the other set \cite{Wong19}. In the next two rows, with the adjacency quantum walk, a single value of $\gamma$ causes the system to evolve from $\ket{s}$ to a mixture of marked and unmarked vertices in both partite sets with a success probability (probability of measuring the position of the walker at a marked vertex) of $1/2 + \sqrt{N_1 N_2}/N$; this same value of $\gamma$ causes the system to evolve from another state $\ket{s_A}$ (this is called $\ket{\sigma}$ in \cite{Wong19}) to a superposition over the marked vertices with certainty \cite{Wong19}. In this research, we show that the signless Laplacian exhibits a combination of the traits of the Laplacian and adjacency quantum walks, as summarized in the last four rows of the table. For some value of $\gamma$, the system partially evolves from the uniform state $\ket{s}$ to the marked vertices in one partite set, and it perfectly evolves a different initial state $\ket{s_Q}$ to the marked vertices. A different value of $\gamma$ partially and perfectly evolves these initial states to the marked vertices in the other partite set. All this assumes that $N_1 \not\approx N_2$ so that the graph is not approximately regular. If $N_1 \approx N_2$, as quantified in \cite{Wong19} as $N_1$ and $N_2$ having a difference that scales less than $\sqrt{N}$, then the graph is approximately regular, and the Laplacian, adjacency, and signless Laplacians all evolve identically (up to a global, ignorable phase), and so they all behave like the adjacency quantum walk in \tref{table:summary}. Thus, we take $N_1 \not\approx N_2$ for the rest of the paper, and from \cite{Wong19}, this is quantified by $N_1$ and $N_2$ having a difference that scales at least as $\sqrt{N}$. The definitions of the three initial states, $\ket{s}$, $\ket{s_A}$, and $\ket{s_Q}$, are summarized in \tref{table:initial}, and these will be described more fully in \sref{section:better-initial}.

In the next section, we review spin networks and show how the Laplacian, adjacency, and signless Laplacian quantum walks arise depending on the coupling between adjacent spins. Then, in \sref{section:search-algorithm}, we detail the search algorithm and give numerical examples of its evolution. In \sref{section:analysis}, we analytically prove the behavior of the signless Laplacian search algorithm, showing that the uniform state partially evolves to the marked vertices in one partite set, depending on the jumping rate. In \sref{section:better-initial}, we show that another initial state completely evolves to the marked vertices in one partite set, again depending on the jumping rate. In \sref{section:runtimes}, we show that for some parameter regimes, the signless Laplacian quantum walk searches more quickly than the Laplacian and adjacency quantum walks. We conclude in \sref{section:conclusion}.


\section{\label{section:spin-network}Heisenberg Spin Networks and Quantum Walks}

\begin{figure}
    \includegraphics{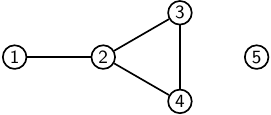}
    \caption{\label{fig:graph-XYZ}A graph of $N = 5$ vertices and $m = 4$ edges.}
\end{figure}

In this section, we review how the Laplacian and adjacency quantum walks can arise in spin networks in the single excitation subspace \cite{Bose2009}, and we additionally show that the signless Laplacian quantum walk can arise as well. A spin network is a model of quantum mechanical spins and is used to study the properties of magnetic systems in statistical mechanics, such as phase transitions and critical points. Each spin can be modeled as the vertex of a graph, and interactions between the spins can be modeled as the edges of the graph. In the Heisenberg (XYZ) model \cite{Heisenberg1928} of such a spin network with no external magnetic field, the Hamiltonian sums over all pairs of adjacent vertices and applies Pauli matrices to the pairs:
\begin{equation}
    \label{eq:Heisenberg}
    H = -\frac{1}{2} \sum_{i \sim j} \left( J_x X_i X_j + J_y Y_i Y_j + J_z Z_i Z_j \right),
\end{equation}
where $X_i$, $Y_i$, and $Z_j$ are the Pauli matrices applied to the $i$th spin in the network, and the $J$'s are coupling constants. That is,
\[ X_i = \underbrace{I_2 \otimes \dots \otimes I_2}_\text{$i-1$ times} \otimes X \otimes \underbrace{I_2 \otimes \dots \otimes I_2}_\text{$N-i$ times}, \]
and similarly for $Y_i$ and $Z_i$. Then, in the Hamiltonian \eqref{eq:Heisenberg}, $X_i X_j$ applies the Pauli-X matrix to the $i$th and $j$th spins, while leaving the other spins unchanged, and $Y_i Y_j$ and $Z_i Z_j$ behave similarly. For example, the graph in \fref{fig:graph-XYZ} has five vertices, and there are four edges or pairs of adjacent vertices---1 and 2, 2 and 3, 2 and 4, and 3 and 4---so the Hamiltonian is
\begin{align*}
    H = -\frac{1}{2} \big[ 
        & \left( J_x X_1 X_2 + J_y Y_1 Y_2 + J_z Z_1 Z_2 \right) \\
        &+ \left( J_x X_2 X_3 + J_y Y_2 Y_3 + J_z Z_2 Z_3 \right) \\
        &+ \left( J_x X_2 X_4 + J_y Y_2 Y_4 + J_z Z_2 Z_4 \right) \\
        &+ \left( J_x X_3 X_4 + J_y Y_3 Y_4 + J_z Z_3 Z_4 \right) \big].
\end{align*}

In terms of the eigenvectors of $Z$, which we write as $\ket{0}_z$ and $\ket{1}_z$ with respective eigenvalues $1$ and $-1$, the Pauli matrices act by
\begin{equation}
\label{eq:Pauli}
\begin{aligned}
    X \ket{0}_z &= \ket{1}_z, \\
    X \ket{1}_z &= \ket{0}_z, \\
    Y \ket{0}_z &= i \ket{1}_z, \\
    Y \ket{1}_z &= -i \ket{0}_z, \\
    Z \ket{0}_z &= \ket{0}_z, \\
    Z \ket{1}_z &= -\ket{1}_z.
\end{aligned}
\end{equation}
Then, writing the eigenstates of $Z$ as basis vectors
\[ \ket{0}_z = \begin{pmatrix}
	1 \\
	0 \\
\end{pmatrix}, \quad
\ket{1}_z = \begin{pmatrix}
	0 \\
	1 \\
\end{pmatrix}, \]
the Pauli matrices are
\[ X = \begin{pmatrix}
	0 & 1 \\
	1 & 0 \\
\end{pmatrix}, \quad
Y = \begin{pmatrix}
	0 & -i \\
	i & 0 \\
\end{pmatrix}, \quad
Z = \begin{pmatrix}
	1 & 0 \\
	0 & -1 \\
\end{pmatrix}. \]

Now, consider the subspace of a single excitation in $Z$, in which each basis state has a single ``1'' eigenstate of $Z$, with the rest ``0'' eigenstates of $Z$:
\begin{gather*}
    \ket{1} = \ket{1}_z \otimes \ket{0}_z \otimes \ket{0}_z \otimes \dots \otimes \ket{0}_z = \ket{100\cdots0}_z, \\
    \ket{2} = \ket{0}_z \otimes \ket{1}_z \otimes \ket{0}_z \otimes \dots \otimes \ket{0}_z = \ket{010\cdots0}_z, \\
    \vdots \\
    \ket{N} = \ket{0}_z \otimes \ket{0}_z \otimes \ket{0}_z \otimes \dots \otimes \ket{1}_z = \ket{000\cdots1}_z.
\end{gather*}
For example, for the graph in \fref{fig:graph-XYZ}, the single-excitation subspace is spanned by
\begin{align*}
    \ket{1} &= \ket{10000}_z, \\
    \ket{2} &= \ket{01000}_z, \\
    \ket{3} &= \ket{00100}_z, \\
    \ket{4} &= \ket{00010}_z, \\
    \ket{5} &= \ket{00001}_z.
\end{align*}
As we will show in the next subsections, the Heisenberg model constitutes a quantum walk on $\ket{1}, \ket{2}, \dots, \ket{N}$ when $J_x = J_y$, and depending on the value of $J_z$, we get either the adjacency quantum walk, Laplacian quantum walk, or signless Laplacian quantum walk.


\subsection{Adjacency Quantum Walk}

When $J_x = J_y = \gamma$ and $J_z = 0$, the Hamiltonian \eqref{eq:Heisenberg} becomes
\[ H_{XY} = -\frac{\gamma}{2} \sum_{i \sim j} \left( X_i X_j + Y_i Y_j \right). \]
This is also known as the XY model. For example, for the graph in \fref{fig:graph-XYZ},
\begin{equation}
\label{eq:H-XY-Graph}
\begin{aligned}
    H_{XY} = -\frac{\gamma}{2} \big[ 
        &\left( X_1 X_2 + Y_1 Y_2 \right) + \left( X_2 X_3 + Y_2 Y_3 \right) \\
        &+\left( X_2 X_4 + Y_2 Y_4 \right) + \left( X_3 X_4 + Y_3 Y_4 \right) \big].
\end{aligned}
\end{equation}

Since $X_i X_j$ only acts on vertices $i$ and $j$, we can focus on the $Z$ basis states corresponding to vertices $i$ and $j$ and ignore the rest. Then, $X_i X_j$ is equivalent to $X \otimes X$ acting on $\ket{0}_z \ket{0}_z$ if the excitation is neither at vertices $i$ nor $j$, $\ket{0}_z \ket{1}_z$ if the excitation is at vertex $j$, and $\ket{1}_z \ket{0}_z$ if the excitation is at vertex $i$. It cannot act on $\ket{1}_z \ket{1}_z$ since there is only one excitation. Working out each of the cases using \eqref{eq:Pauli},
\begin{align*}
    (X \otimes X) \ket{0}_z \ket{0}_z &= \ket{1}_z \ket{1}_z, \\
    (X \otimes X) \ket{0}_z \ket{1}_z &= \ket{1}_z \ket{0}_z, \\
    (X \otimes X) \ket{1}_z \ket{0}_z &= \ket{0}_z \ket{1}_z.
\end{align*}
Similarly, $Y_i Y_j$ is equivalent to $Y \otimes Y$ acting on the three possible single-excitation $Z$ basis states for vertices $i$ and $j$, and using \eqref{eq:Pauli}, we get
\begin{align*}
    (Y \otimes Y) \ket{0}_z \ket{0}_z &= \left(i \ket{1}_z \right) \left( i \ket{1}_z \right) = -\ket{1}_z \ket{1}_z, \\
    (Y \otimes Y) \ket{0}_z \ket{1}_z &= \left(i \ket{1}_z \right) \left( -i \ket{0}_z \right) = \ket{1}_z \ket{0}_z, \\
    (Y \otimes Y) \ket{1}_z \ket{0}_z &= \left(-i \ket{0}_z \right) \left( i \ket{1}_z \right) = \ket{0}_z \ket{1}_z.
\end{align*}
Combining these, $X_i X_j + Y_i Y_j$ acts by
\begin{align*}
    \left( X \otimes X + Y \otimes Y \right) \ket{0}_z \ket{0}_z &= 0, \\
    \left( X \otimes X + Y \otimes Y \right) \ket{0}_z \ket{1}_z &= 2 \ket{1}_z \ket{0}_z, \\
    \left( X \otimes X + Y \otimes Y \right) \ket{1}_z \ket{0}_z &= 2 \ket{0}_z \ket{1}_z.
\end{align*}
Thus, $X_i X_j + Y_i Y_j$ has the effect of moving the excitation between adjacent vertices, and the Hamiltonian is equal to
\begin{equation}
    \label{eq:H-XY-A}
    H_{XY} = -\gamma A,
\end{equation}
which is an adjacency quantum walk.

For example, to show \eqref{eq:H-XY-A} explicitly for the graph in \fref{fig:graph-XYZ}, $H_{XY}$ \eqref{eq:H-XY-Graph} acts on each of the five single-excitation basis states as
\begin{align*}
    H_{XY} \ket{1}
        &= H_{XY} \ket{10000}_z \\
        &= -\frac{\gamma}{2} \left[ 2\ket{01000}_z + 0 + 0 + 0 \right] \\
        &= -\gamma \ket{2}, \displaybreak[0] \\
    H_{XY} \ket{2}
        &= H_{XY} \ket{01000}_z \\
        &= -\frac{\gamma}{2} \left[ 2\ket{10000}_z + 2\ket{00100}_z + 2\ket{00010}_z + 0 \right] \\
        &= -\gamma \left[ \ket{1} + \ket{3} + \ket{4} \right], \displaybreak[0] \\
    H_{XY} \ket{3}
        &= H_{XY} \ket{00100}_z \\
        &= -\frac{\gamma}{2} \left[ 0 + 2\ket{01000}_z + 0 + 2\ket{00010}_z \right] \\
        &= -\gamma \left[ \ket{2} + \ket{4} \right], \displaybreak[0] \\
    H_{XY} \ket{4}
        &= H_{XY} \ket{00010}_z \\
        &= -\frac{\gamma}{2} \left[ 0 + 2\ket{01000}_z + 0 + 2\ket{00100}_z \right] \\
        &= -\gamma \left[ \ket{2} + \ket{3} \right], \displaybreak[0] \\
    H_{XY} \ket{5} \\
        &= H_{XY} \ket{00001}_z \\
        &= -\frac{\gamma}{2} \left[ 0 + 0 + 0 + 0 \right] \\
        &= 0.
\end{align*}
Then,
\begin{align*}
    H_{XY}
    &= -\gamma \begin{pmatrix}
        0 & 1 & 0 & 0 & 0 \\
        1 & 0 & 1 & 1 & 0 \\
        0 & 1 & 0 & 1 & 0 \\
        0 & 1 & 1 & 0 & 0 \\
        0 & 0 & 0 & 0 & 0 \\
    \end{pmatrix} \\
    &= -\gamma A,
\end{align*}
which demonstrates \eqref{eq:H-XY-A}.


\subsection{Laplacian Quantum Walk}

When $J_x = J_y = J_z = \gamma$, the Hamiltonian \eqref{eq:Heisenberg} becomes
\begin{align*}
    H_{XYZ}
        &= -\frac{\gamma}{2} \sum_{i \sim j} \left( X_i X_j + Y_i Y_j + Z_i Z_j \right) \\
        &= H_{XY} + H_Z,
\end{align*}
where $H_{XY} = -\gamma A$ \eqref{eq:H-XY-A}, and
\[ H_z = -\frac{\gamma}{2} \sum_{i \sim j} Z_i Z_j. \]
For example, for the graph in \fref{fig:graph-XYZ},
\begin{equation}
    \label{eq:H-Z-Graph}
    H_z = -\frac{\gamma}{2} \big( Z_1 Z_2 + Z_2 Z_3 + Z_2 Z_4 + Z_3 Z_4 \big).
\end{equation}
So, we need to determine how $Z_i Z_j$ acts. Since it is equivalent to $Z \otimes Z$ on the three possible single-excitation states for vertices $i$ and $j$,
\begin{align*}
    (Z \otimes Z) \ket{0}_z \ket{0}_z &= \ket{0}_z \ket{0}_z, \\
    (Z \otimes Z) \ket{0}_z \ket{1}_z &= \ket{0}_z \left( - \ket{1}_z \right) = -\ket{0}_z \ket{1}_z, \\
    (Z \otimes Z) \ket{1}_z \ket{0}_z &= \left(- \ket{1}_z \right) \ket{0}_z = -\ket{1}_z \ket{0}_z.
\end{align*}
So, when neither vertex $i$ nor $j$ has an excitation, $Z_i Z_j$ does nothing, while if one of them has the excitation, it flips the sign of the state.

$H_Z$ is a collection of $Z_i Z_j$'s, and to see its behavior, let us work out how \eqref{eq:H-Z-Graph} acts on the single-excitation basis states for the graph in \fref{fig:bipartite}
\begin{align*}
    H_Z \ket{1}
        &= H_Z \ket{10000}_z \\
        &= -\frac{\gamma}{2} \left( -1 + 1 + 1 + 1 \right) \ket{10000}_z \\
        &= -\frac{\gamma}{2} \left[ 4 - 2(1) \right] \ket{1}, \displaybreak[0] \\
    H_Z \ket{2}
        &= H_Z \ket{01000}_z \\
        &= -\frac{\gamma}{2} \left( -1 - 1 - 1 + 1 \right) \ket{01000}_z \\
        &= -\frac{\gamma}{2} \left[ 4 - 2(3) \right] \ket{2}, \displaybreak[0] \\
    H_Z \ket{3}
        &= H_Z \ket{00100}_z \\
        &= -\frac{\gamma}{2} \left( 1 - 1 + 1 - 1 \right) \ket{00100}_z \\
        &= -\frac{\gamma}{2} \left[ 4 - 2(2) \right] \ket{3}, \displaybreak[0] \\
    H_Z \ket{4}
        &= H_Z \ket{0001}_z \\
        &= -\frac{\gamma}{2} \left( 1 + 1 - 1 - 1 \right) \ket{00010}_z \\
        &= -\frac{\gamma}{2} \left[ 4 - 2(2) \right] \ket{4}, \displaybreak[0] \\
    H_Z \ket{5}
        &= H_Z \ket{00001}_z \\
        &= -\frac{\gamma}{2} \left( 1 + 1 + 1 + 1 \right) \ket{00001}_z \\
        &= -\frac{\gamma}{2} \left[ 4 - 2(0) \right] \ket{5}.
\end{align*}
In each calculation, the four $\pm 1$ terms arise because there are four edges, and so there are four $Z_i Z_j$ terms in $H_z$. If a vertex has no neighbors, all of the terms are $+1$, but for each additional neighbor, one of the $+1$'s gets turned into a $-1$, which is a difference of $2$. Thus, the term in square brackets is the number of edges minus twice the number of neighbors. Then,
\[ H_Z = -\frac{\gamma}{2} \left( 4 I_5 - 2 D \right), \]
where
\[ D = \begin{pmatrix}
    1 & 0 & 0 & 0 & 0 \\
    0 & 3 & 0 & 0 & 0 \\
    0 & 0 & 2 & 0 & 0 \\
    0 & 0 & 0 & 2 & 0 \\
    0 & 0 & 0 & 0 & 0 \\
\end{pmatrix} \]
is the degree matrix of the graph.

In general, for a graph with $N$ vertices and $m$ edges,
\begin{align*}
    H_z 
        &= -\frac{\gamma}{2} \left( m I_N - 2 D \right) \\
        &= \gamma D - \frac{\gamma m}{2} I_N.
\end{align*}
A multiple of the identity matrix in the Hamiltonian contributes a global, unobservable phase, so it can be dropped, which can also be interpreted as a rezeroing of energy. Then,
\begin{equation}
    \label{eq:H-Z-D}
    H_z = \gamma D.
\end{equation}
Thus,
\begin{align*}
    H_{XYZ} 
        &= H_{XY} + H_Z \\
        &= -\gamma A + \gamma D \\
        &= -\gamma(A - D) \\
        &= -\gamma L,
\end{align*}
which is the Laplacian quantum walk.


\subsection{Signless Laplacian Quantum Walk}

Layered ferromagnetic materials with antiferromagnetic interactions between the layers can have positive coupling constants $J_x$ and $J_y$ with negative $J_z$ \cite{deJongh1972,Lai1999}. Motivated by such systems, when $J_x = J_y = \gamma$ and $J_z = -\gamma$, the Hamiltonian \eqref{eq:Heisenberg} becomes
\begin{align*}
    H_-
        &= -\frac{\gamma}{2} \sum_{i \sim j} \left( X_i X_j + Y_i Y_j - Z_i Z_j \right) \\
        &= H_{XY} - H_Z \\
        &= -\gamma A - \gamma D \\
        &= -\gamma (A+D) \\
        &= -\gamma Q,
\end{align*}
where in the third line, we used \eqref{eq:H-XY-A} and \eqref{eq:H-Z-D}. This is the signless Laplacian quantum walk.


\section{\label{section:search-algorithm}Search Algorithm}

In the quantum walk search algorithm \cite{CG2004}, a potential energy term is added to the Hamiltonian, which acts as an oracle marking the vertices to be found, and so for the signless Laplacian, the search Hamiltonian is
\begin{equation}
    \label{eq:H-search}
    H = -\gamma Q - \sum_{i\text{ marked}} \ketbra{i}{i}.
\end{equation}
In typical quantum search algorithms \cite{CG2004,Novo2015}, the initial state is a uniform superposition over all $N$ vertices, which we call $\ket{s}$:
\begin{equation}
    \label{eq:s}
    \ket{s} = \frac{1}{\sqrt{N}} \sum_{i=1}^N \ket{i}.
\end{equation}
This reflects our initial lack of knowledge of which vertex is marked; since every vertex is equally likely to be marked, each starts with an initial probability of $1/N$. Now, this uniform state evolves by Schr\"odinger's equation \eqref{eq:Schrodinger} with Hamiltonian \eqref{eq:H-search}, and since the Hamiltonian is time-independent, the state at time $t$ is
\begin{equation}
    \label{eq:psi(t)}
    \ket{\psi(t)} = e^{-iHt} \ket{s}.
\end{equation}

\begin{figure}
\begin{center}
    \subfloat[] {
        \label{fig:s-prob-time-gamma-a}
        \includegraphics{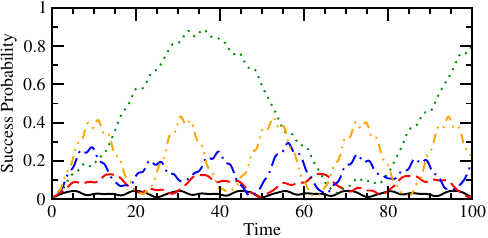}
    }

    \subfloat[] {
        \label{fig:s-prob-time-gamma-b}
        \includegraphics{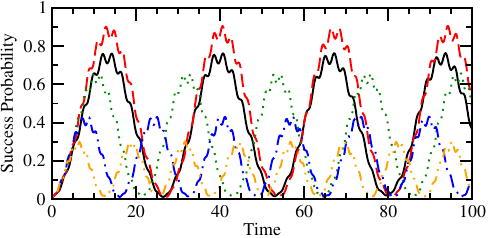}
    }
    \caption{\label{fig:s-prob-time-gamma}Success probability versus time for search on a complete bipartite graph of $N_1 = 512$ and $N_2 = 256$ vertices with $k_1 = 3$ and $k_2 = 5$ marked vertices, starting with the uniform state $\ket{s}$ and using a signless Laplacian quantum walk with various jumping rates. (a) The solid black curve is $\gamma = 0.001$, the dashed red curve is $\gamma = 0.0015$, the dotted green curve is $\gamma = 0.002$, the dot-dashed blue curve is $\gamma = 0.0025$, and the dot-dot-dashed orange curve is $\gamma = 0.003$. (b) The solid black curve is $\gamma = 0.0035$, the dashed red curve is $\gamma = 0.004$, the dotted green curve is $\gamma = 0.0045$, the dot-dashed blue curve is $\gamma = 0.005$, and the dot-dot-dashed orange curve is $\gamma = 0.0055$.}
\end{center}
\end{figure}

The performance of the algorithm depends on the jumping rate $\gamma$. To illustrate this, in \fref{fig:s-prob-time-gamma}, we have plotted the success probability versus time with various values of $\gamma$. That is, we are plotting
\[ p(t) = \sum_{i\text{ marked}} \left| \braket{i}{\psi(t)} \right|^2. \]
Starting with \fref{fig:s-prob-time-gamma-a}, the solid black curve depicts $\gamma = 0.001$, and the success probability stays near its initial value of $8/768 \approx 0.01$. Increasing the jumping rate, when $\gamma = 0.0015$, the dashed red curve indicates that the success probability now rises a little. Continuing, when $\gamma = 0.002$, the dotted green curve reveals that the success probability increases significantly, and measuring the location of the walker finds it at a marked vertex with high probability. Further still, when $\gamma = 0.0025$, the dot-dashed blue curve shows that the success probability does not build up as well. When $\gamma = 0.003$, the dot-dot-dashed orange curve reveals that the success probability is building up again. Continuing in \fref{fig:s-prob-time-gamma-b}, when $\gamma = 0.0035$, the solid black curve shows that the success probability is increasing still. The dashed red curve is $\gamma = 0.004$, and the success probability has reached a high peak. Increasing $\gamma$ further to $0.0045$, $0.005$, and $0.0055$, the dotted green, dot-dashed blue, and dot-dot-dashed orange curves show that the success probability gets worse and worse. Altogether, these plots indicate that there are two values of $\gamma$, called critical values, for which the success probability is maximized.

\begin{figure}
\begin{center}
    \subfloat[] {
        \label{fig:overlaps-s}
        \includegraphics{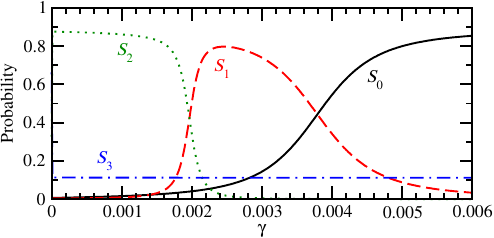}
    }

    \subfloat[] {
        \label{fig:overlaps-a}
        \includegraphics{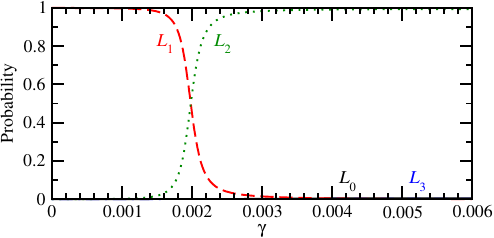}
    }

    \subfloat[] {
        \label{fig:overlaps-b}
        \includegraphics{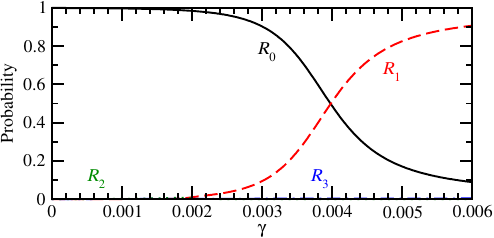}
    }
    \caption{\label{fig:overlaps}For the complete bipartite graph of $N_1 = 512$ and $N_2 = 256$ vertices with $k_1 = 3$ and $k_2 = 5$ marked vertices, the norm-square of the overlaps of the eigenvectors of the search Hamiltonian with (a) the uniform state, (b) the marked vertices in the left partite set, and (c) the marked vertices in the right partite set. In all three subfigures, the solid black curve is the ground state $\ket{\psi_0}$, the dashed red curve is the first excited state $\ket{\psi_1}$, the dotted green curve is the second excited state $\ket{\psi_2}$, and the dot-dashed blue curve is the third excited state $\ket{\psi_3}$.}
\end{center}
\end{figure}

One way to visualize the two critical values of $\gamma$ is by plotting the norm-square of the inner product of the eigenvectors $\ket{\psi_0}, \ket{\psi_1}, \dots$ of the search Hamiltonian \eqref{eq:H-search} with the initial state $\ket{s}$, with the $k_1$ marked vertices in the left partite set, and with the $k_2$ marked vertices in the right partite set, as functions of $\gamma$, i.e., the quantities
\begin{align*}
    S_n &= \left| \braket{s}{\psi_n} \right|^2, \displaybreak[0] \\
    L_n &= \sum_{i\text{ left marked}} \left| \braket{i}{\psi_n} \right|^2, \displaybreak[0] \\
    R_n &= \sum_{i\text{ right marked}} \left| \braket{i}{\psi_n} \right|^2,
\end{align*}
with $n = 0, 1, \dots$ as functions of the jumping rate. This is shown in \fref{fig:overlaps} for the first four eigenvectors of the search Hamiltonian, using the same parameters for the complete bipartite graph as in \fref{fig:s-prob-time-gamma}. Let us examine these figures for five regions/values of increasing $\gamma$: when it is at the far left, at the first crossing around $\gamma = 0.002$, in the middle, at the second crossing around $\gamma = 0.004$, and at the far right:
\begin{enumerate}
    \item When the jumping rate is small, say $\gamma = 0.001$ or $\gamma = 0.0015$, then \fref{fig:overlaps-s} indicates that the initial state $\ket{s}$ is primarily in the second excited state $\ket{\psi_2}$, and secondarily in the third excited state $\ket{\psi_3}$. Then, in \fref{fig:overlaps-a} and \fref{fig:overlaps-b}, when $\gamma = 0.001$, neither $\ket{\psi_2}$ nor $\ket{\psi_3}$ have a substantial overlap with the marked vertices in either the left or right partite sets. Then, the system does not evolve to a state with a substantial success probability, as we saw in the solid black and dashed red curves in \fref{fig:s-prob-time-gamma-a}.
    
    \item In \fref{fig:overlaps-s}, the dashed red and dotted green curves cross around $\gamma = 0.002$. At this jumping rate, $\ket{s}$ is primarily in a combination of the first excited state $\ket{\psi_1}$ and second excited state $\ket{\psi_2}$, with a smaller contribution in $\ket{\psi_3}$. Then, in \fref{fig:overlaps-a} and \fref{fig:overlaps-b}, $\ket{\psi_1}$ and $\ket{\psi_2}$ both have significant overlaps with the marked vertices in the left partite set, but an insignificant overlap with the marked vertices in the right partite set, while $\ket{\psi_3}$ has insignificant overlaps with both. Then, we expect that the portion of the initial state in $\ket{\psi_1}$ and $\ket{\psi_2}$ will evolve to the marked vertices in the left partite set, resulting in a high success probability, as we saw in the dotted green curve in \fref{fig:s-prob-time-gamma-a}.
    
    \item When the jumping rate is in the middle, say $\gamma = 0.0025$ or $\gamma = 0.003$, then \fref{fig:overlaps-s} indicates that the initial state $\ket{s}$ is primarily in $\ket{\psi_1}$, which smaller contributions in $\ket{\psi_0}$ and $\ket{\psi_3}$. With this jumping rate, \fref{fig:overlaps-a} and \fref{fig:overlaps-b} indicate that $\ket{\psi_1}$ and $\ket{\psi_0}$ have a little overlap with the marked vertices in the left and right partite sets, respectively, while $\ket{\psi_3}$ has almost no overlap with them. Then, the system should only evolve a little to the marked vertices, and this is consistent with the dot-dashed blue and dot-dot-dashed orange curves is \fref{fig:s-prob-time-gamma-a}.
    
    \item As $\gamma$ gets larger, we approach another crossing in the overlap of $\ket{s}$ with the eigenvectors of the search Hamiltonian. From \fref{fig:overlaps-s}, this occurs around $\gamma = 0.0035$, $0.004$, and $0.0045$, where $\ket{s}$ has significant contributions in $\ket{\psi_0}$ and $\ket{\psi_1}$, and a minor contribution in $\ket{\psi_3}$, and from \fref{fig:overlaps-b}, these eigenvectors also have nontrivial overlaps with the marked vertices in the right partite set. As a result, evolution results in a high success probability, consistent with the solid black, dashed red, and dotted green curves in \fref{fig:s-prob-time-gamma-b}, with the dashed red $\gamma = 0.004$ curve performing the best among these.
    
    \item When the jumping rate is large, say $\gamma = 0.005$ and $0.0055$, then \fref{fig:overlaps-a} indicates that the initial state $\ket{s}$ is increasingly in $\ket{\psi_0}$, and from \fref{fig:overlaps-b}, $\ket{\psi_0}$ has decreasing overlap with the marked vertices in the right partite set. As a result, the success probabability should get worse as $\gamma$ gets larger, as we saw in the dot-dashed blue and dot-dot-dashed orange curves in \fref{fig:s-prob-time-gamma-b}.
\end{enumerate}
For large $N_1$ and $N_2$, these regions are even more sharply delineated. That is, the system evolves significantly at the crossings, which were cases 2 and 4 above, and is relatively unchanged in the other three cases (cases 1, 3, and 5).


\section{\label{section:analysis}Analytical Proof}

In this section, we analytically prove the behavior of the search algorithm, and in doing so, we will find formulas for the critical values of $\gamma$, the runtime, and the success probability. The general idea is to express the initial state \eqref{eq:s} as a linear combination of the eigenvectors $\{ \ket{\psi_i} \}$ of the search Hamiltonian \eqref{eq:H-search}:
\[ \ket{s} = c_1 \ket{\psi_1} + c_2 \ket{\psi_2} + \dots + c_N \ket{\psi_N}, \]
for some constants $\{ c_i \}$. Then, since the system evolves according to \eqref{eq:psi(t)}, the state at time $t$ is obtained by multiplying each eigenvector by $e^{-i E_i t}$, where $E_i$ is the eigenvalue of the eigenvector $\ket{\psi_i}$:
\begin{align*}
    \ket{\psi(t)} 
        &= c_1 e^{-i E_1 t} \ket{\psi_1} + c_2 e^{-i E_2 t} \ket{\psi_2} + \dots \\
        &\quad + c_N e^{-i E_N t} \ket{\psi_N}.
\end{align*}

In general, $\ket{s}$ is a vector of length $N$, and the search Hamiltonian is an $N \times N$ matrix. For the complete bipartite graph, however, the system evolves in a 4-dimensional (4D) subspace spanned by $\{\ket{a},\ket{b},\ket{c},\ket{d}\}$, where $\ket{a}$ is the uniform superposition of marked vertices in the left partite set, $\ket{b}$ is the uniform superposition of marked vertices in the right partite set, $\ket{c}$ is the uniform superposition of unmarked vertices in the left partite set, and $\ket{d}$ is the uniform superposition of unmarked vertices in the right partite set, i.e.,
\begin{align*}
    \ket{a} &= \frac{1}{\sqrt{k_1}} \sum_{i\text{ left marked}} \ket{i}, \displaybreak[0] \\
    \ket{b} &= \frac{1}{\sqrt{k_2}} \sum_{i\text{ right marked}} \ket{i}, \displaybreak[0] \\
    \ket{c} &= \frac{1}{\sqrt{N_1 - k_1}} \sum_{i\text{ left unmarked}} \ket{i}, \displaybreak[0] \\
    \ket{d} &= \frac{1}{\sqrt{N_2 - k_2}} \sum_{i\text{ right unmarked}} \ket{i}.
\end{align*}
\Fref{fig:bipartite} includes labels indicating whether each vertex belongs to $\ket{a}$, $\ket{b}$, $\ket{c}$, or $\ket{d}$. In this 4D basis, the initial uniform superposition state \eqref{eq:s} is
\begin{equation}
    \label{eq:s-4D}
    \ket{s} = \frac{1}{\sqrt{N}} \begin{pmatrix}
        \sqrt{k_1} \\
        \sqrt{k_2} \\
        \sqrt{N_1 - k_1} \\
        \sqrt{N_2 - k_2} \\
    \end{pmatrix}.
\end{equation}
The adjacency matrix is
\[ A = \begin{pmatrix}
		0 & \sqrt{k_1 k_2} & 0 & \sqrt{k_1 N_{k2}} \\
		\sqrt{k_1 k_2} & 0 & \sqrt{k_2 N_{k1}} & 0 \\
		0 & \sqrt{k_2 N_{k1}} & 0 & \sqrt{N_{k1} N_{k2}} \\
		\sqrt{k_1 N_{k2}} & 0 & \sqrt{N_{k1} N_{k2}} & 0 \\
\end{pmatrix}, \]
where $N_{ki} = N_i - k_i$. For example, the first item of the last row, $\sqrt{k_1 N_{k2}} = \sqrt{k_1(N_2 - k_2)}$ is obtained from the $k_1$ vertices of type $a$ multiplied by $\sqrt{N_2 - k_2}/\sqrt{k_1}$ to convert between the normalizations of the $\ket{a}$ and $\ket{d}$ basis states. The degree matrix is $D = \text{diag}(N_2, N_1, N_2, N_1)$, and the Hamiltonian oracle is $-\sum_{i\text{ marked}} \ketbra{i}{i} = \text{diag}(-1,-1,0,0)$. Putting these together, the search Hamiltonian \eqref{eq:H-search} is
\begin{equation}
	\label{eq:H-search-4D}
	H = -\gamma \begin{pmatrix}
		\frac{1}{\gamma} + N_2 & \sqrt{k_1 k_2} & 0 & \sqrt{k_1 N_{k2}} \\
		\sqrt{k_1 k_2} & \frac{1}{\gamma} + N_1 & \sqrt{k_2 N_{k1}} & 0 \\
		0 & \sqrt{k_2 N_{k1}} & N_2 & \sqrt{N_{k1} N_{k2}} \\
		\sqrt{k_1 N_{k2}} & 0 & \sqrt{N_{k1} N_{k2}} & N_1 \\
	\end{pmatrix}.
\end{equation}

Now, we can find the eigenvectors and eigenvalues of this $4 \times 4$ matrix \eqref{eq:H-search-4D}, express $\ket{s}$ as a linear combination of the eigenvectors, and then multiply each eigenvector by $e^{-iE_i t}$ to find the state at time $t$. Doing this for large $N_1$ and $N_2$, the dominant terms in \eqref{eq:H-search-4D} are
\[ H^{(0)} = -\gamma \begin{pmatrix}
	\frac{1}{\gamma} + N_2 & 0 & 0 & 0 \\
	0 & \frac{1}{\gamma} + N_1 & 0 & 0 \\
	0 & 0 & N_2 & \sqrt{N_1 N_2} \\
	0 & 0 & \sqrt{N_1 N_2} & N_1 \\
\end{pmatrix}. \]
The eigenvectors and eigenvalues of this are
\begin{equation}
    \label{eq:eigensystem-H0}
    \begin{aligned}
        & \ket{a}, && -1 - \gamma N_2, \\
        & \ket{b}, && -1 - \gamma N_1, \\
        & \ket{u} = \frac{1}{\sqrt{N}} \left( \sqrt{N_2} \ket{c} + \sqrt{N_1} \ket{d} \right), && -\gamma N, \\
        & \ket{v} = \frac{1}{\sqrt{N}} \left( \sqrt{N_1} \ket{c} - \sqrt{N_2} \ket{d} \right), && 0.
    \end{aligned}
\end{equation}
Next, in the following subsections, we continue the analysis for three cases: 
\begin{enumerate}[label=(\roman*)]
    \item   When none of the above eigenvectors are degenerate. That is, $\gamma \ne 1/N_1$ and $\gamma \ne 1/N_2$, which corresponds to regions 1, 3, and 5 at the end of \sref{section:search-algorithm}.
    \item   When $\ket{a}$ and $\ket{u}$ are degenerate. That is, $\gamma = 1/N_1$, which corresponds to region 2 at the end of \sref{section:search-algorithm}.
    \item   When $\ket{b}$ and $\ket{u}$ are degenerate. That is, $\gamma = 1/N_2$, which corresponds to region 4 at the end of \sref{section:search-algorithm}.
\end{enumerate}


\subsection{Nondegenerate Case}

For case (i), we express the initial uniform state \eqref{eq:s-4D} for large $N_1$ and $N_2$, and then express the state as a linear combination of the eigenvectors of $H^{(0)}$:
\begin{align*}
    \ket{s} 
        &\approx \frac{1}{\sqrt{N}} \left( \sqrt{N_1} \ket{c} + \sqrt{N_2} \ket{d} \right) \\
        &= \frac{2\sqrt{N_1 N_2}}{N} \ket{u} + \frac{N_1 - N_2}{N} \ket{v}.
\end{align*}
Then, the state at time $t$ is
\begin{align*}
    \ket{\psi(t)} 
        &\approx \frac{2\sqrt{N_1 N_2}}{N} e^{i\gamma Nt} \ket{u} \\
        &\quad + \frac{N_1 - N_2}{N} e^{0} \ket{v} \\
        &= \frac{2\sqrt{N_1 N_2}}{N} e^{i\gamma Nt} \frac{1}{\sqrt{N}} \left( \sqrt{N_2} \ket{c} + \sqrt{N_1} \ket{d} \right) \\
        &\quad + \frac{N_1 - N_2}{N} \frac{1}{\sqrt{N}} \left( \sqrt{N_1} \ket{c} - \sqrt{N_2} \ket{d} \right).
\end{align*}
Since this has no amplitude at the marked vertices (i.e., $\braket{a}{\psi(t)} = 0$ and $\braket{b}{\psi(t)} = 0$), the success probability is asymptotically zero for large $N_1$ and $N_2$ for these regions of $\gamma$.


\subsection{First Degenerate Case}

For case (ii), $\gamma = 1/N_1$, and so $\ket{a}$ and $\ket{u}$ are degenerate eigenvectors of $H^{(0)}$ with eigenvalue $-1 - N_2/N_1$. Using degenerate perturbation theory \cite{Griffiths2018}, we can ``lift'' the degeneracy by finding linear combinations of $\ket{a}$ and $\ket{u}$ that are eigenvectors of $H^{(0)}$ plus its first-order corrections
\[ H^{(1)} = -\gamma \begin{pmatrix}
	0 & 0 & 0 & \sqrt{k_1 N_{2}} \\
	0 & 0 & \sqrt{k_2 N_{1}} & 0 \\
	0 & \sqrt{k_2 N_{1}} & 0 & 0 \\
	\sqrt{k_1 N_{2}} & 0 & 0 & 0 \\
\end{pmatrix}, \]
where $H^{(1)}$ are the second most significant terms in the search Hamiltonian \eqref{eq:H-search-4D}. That is, we are solving for linear combinations
\[ \alpha_a \ket{a} + \alpha_u \ket{u} \]
that satisfy the eigenvalue relation
\[ \left( H^{(0)} + H^{(1)} \right) \left( \alpha_a \ket{a} + \alpha_u \ket{u} \right)) = E \left( \alpha_a \ket{a} + \alpha_u \ket{u} \right). \]
In the $\{ \ket{a}, \ket{u} \}$ basis, this eigenvalue relation is
\[ \begin{pmatrix}
    -\frac{N}{N_1} & -\sqrt{\frac{k_1 N_2}{N_1 N}} \\
    -\sqrt{\frac{k_1 N_2}{N_1 N}} & -\frac{N}{N_1} \\
\end{pmatrix} \begin{pmatrix}
    \alpha_a \\
    \alpha_u \\
\end{pmatrix} = E \begin{pmatrix}
    \alpha_a \\
    \alpha_u \\
\end{pmatrix}, \]
which we solve to yield two eigenvectors and eigenvalues of $H^{(0)} + H^{(1)}$:
\[ \begin{pmatrix}
    \alpha_a \\
    \alpha_u \\
\end{pmatrix} = \frac{1}{\sqrt{2}} \begin{pmatrix}
    1 \\
    \pm 1 \\
\end{pmatrix}, \enspace E = -1 - \frac{N_2}{N_1} \mp \sqrt{\frac{k_1 N_2}{N_1 N}}. \]
Combining these with the non-degenerate eigenvectors of $H^{(0)}$ when $\gamma = 1/N_1$ from \eqref{eq:eigensystem-H0}, the four asymptotic eigenvectors of $H$ when $\gamma= 1/N_1$ are
\begin{equation}
    \label{eq:eigensystem-gamma1}
    \begin{aligned}
        &\ket{\psi_0} = \ket{b}, \\
        &\ket{\psi_1} = \frac{1}{\sqrt{2}} \left( \ket{a} + \ket{u} \right), \\
        &\ket{\psi_2} = \frac{1}{\sqrt{2}} \left( \ket{a} - \ket{u} \right), \\
        &\ket{\psi_3} = \ket{v},
    \end{aligned}
\end{equation}
with corresponding eigenvalues
\begin{align*}
    & E_0 = -2, \\
    & E_1 = -1 - \frac{N_2}{N_1} - \sqrt{\frac{k_1 N_2}{N_1 N}}, \\
    & E_2 = -1 - \frac{N_2}{N_1} + \sqrt{\frac{k_1 N_2}{N_1 N}}, \\
    & E_3 = 0.
\end{align*}
Note the ordering of these eigenvalues are from smallest to largest using the values of $N_1$, $N_2$, $k_1$, and $k_2$ from \fref{fig:s-prob-time-gamma} and \fref{fig:overlaps}, but for different parameters, their order might not be correct. This does not, however, affect the results of the analysis.

\begin{figure}
\begin{center}
    \includegraphics{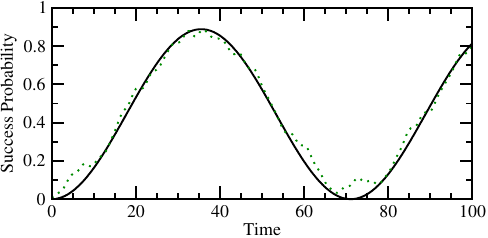}
    \caption{\label{fig:s-prob-time-exact-asymptotic}Success probability versus time for search on a complete bipartite graph of $N_1 = 512$ and $N_2 = 256$ vertices with $k_1 = 3$ and $k_2 = 5$ marked vertices, starting with the uniform state $\ket{s}$ and using a signless Laplacian quantum walk. The dotted green curve is $\gamma = 0.002$ and is copied from \fref{fig:s-prob-time-gamma-a}, and the solid black curve is the asymptotic formula $p_a(t)$ derived in \eqref{eq:s-prob-time}.}
\end{center}
\end{figure}

Now that we have the eigenvectors and eigenvalues of the search Hamiltonian \eqref{eq:H-search-4D} for large $N_1$ and $N_2$ when $\gamma = 1/N_1$, we can follow the general idea of expressing the initial uniform state as a linear combination of them, and then multiply each eigenvector by a complex phase to get the state at time $t$. First, the initial state for large $N$ is
\begin{align*}
    \ket{s} 
        &\approx \frac{1}{\sqrt{N}} \left( \sqrt{N_1} \ket{c} + \sqrt{N_2} \ket{d} \right) \\
        &= \frac{\sqrt{2N_1N_2}}{N} \ket{\psi_1} - \frac{\sqrt{2N_1N_2}}{N} \ket{\psi_2} + \frac{N_1-N_2}{N} \ket{\psi_3}.
\end{align*}
Then, the system evolves to
\begin{widetext}
\begin{align*}
    \ket{\psi(t)}
        &\approx \frac{\sqrt{2N_1N_2}}{N} e^{-iE_1t} \ket{\psi_1} - \frac{\sqrt{2N_1N_2}}{N} e^{-iE_2t} \ket{\psi_2} + \frac{N_1-N_2}{N} e^{-iE_3t} \ket{\psi_3} \displaybreak[0] \\
        &= \frac{\sqrt{2N_1N_2}}{N} e^{-iE_1t} \frac{1}{\sqrt{2}} \left( \ket{a} + \ket{u} \right) - \frac{\sqrt{2N_1N_2}}{N} e^{-iE_2t} \frac{1}{\sqrt{2}} \left( \ket{a} - \ket{u} \right) + \frac{N_1-N_2}{N} e^0 \ket{v} \displaybreak[0] \\
        &= \frac{\sqrt{2N_1N_2}}{N} e^{-iE_1t} \frac{1}{\sqrt{2}} \left[ \ket{a} + \frac{1}{\sqrt{N}} \left( \sqrt{N_2} \ket{c} + \sqrt{N_1} \ket{d} \right) \right] \\
        &\quad - \frac{\sqrt{2N_1N_2}}{N} e^{-iE_2t} \frac{1}{\sqrt{2}} \left[ \ket{a} - \frac{1}{\sqrt{N}} \left( \sqrt{N_2} \ket{c} + \sqrt{N_1} \ket{d} \right) \right] \\
        &\quad + \frac{N_1-N_2}{N} \frac{1}{\sqrt{N}} \left( \sqrt{N_1} \ket{c} - \sqrt{N_2} \ket{d} \right) \displaybreak[0] \\
        &= \frac{\sqrt{N_1N_2}}{N} \left( e^{-iE_1t} - e^{-iE_2t} \right) \ket{a} + \left[ \frac{\sqrt{N_1}N_2}{N^{3/2}} \left( e^{-iE_1t} + e^{-iE_2t} \right) + \frac{\sqrt{N_1}(N_1 - N_2)}{N^{3/2}} \right] \ket{c} \\
        &\quad + \left[ \frac{N_1\sqrt{N_2}}{N^{3/2}} \left( e^{-iE_1t} + e^{-iE_2t} \right) - \frac{\sqrt{N_2}(N_1 - N_2)}{N^{3/2}} \right] \ket{d} \displaybreak[0] \\
        &= \frac{\sqrt{N_1N_2}}{N} e^{i(1+N_2/N_1)t} \left( e^{i\Delta E t/2} - e^{-i\Delta E t/2} \right) \ket{a} \\
        &\quad + \left[ \frac{\sqrt{N_1}N_2}{N^{3/2}} e^{i(1+N_2/N_1)t} \left( e^{i\Delta E t/2} + e^{-i\Delta E t/2} \right) + \frac{\sqrt{N_1}(N_1 - N_2)}{N^{3/2}} \right] \ket{c} \\
        &\quad + \left[ \frac{N_1\sqrt{N_2}}{N^{3/2}} e^{i(1+N_2/N_1)t} \left( e^{i\Delta E t/2} + e^{-i\Delta E t/2} \right) - \frac{\sqrt{N_2}(N_1 - N_2)}{N^{3/2}} \right] \ket{d} \displaybreak[0] \\
        &= \frac{\sqrt{N_1N_2}}{N} e^{i(1+N_2/N_1)t} 2i \sin \left( \frac{\Delta E}{2} t \right) \ket{a} \\
        &\quad + \left[ \frac{\sqrt{N_1}N_2}{N^{3/2}} e^{i(1+N_2/N_1)t} 2 \cos \left( \frac{\Delta E}{2} t \right) + \frac{\sqrt{N_1}(N_1 - N_2)}{N^{3/2}} \right] \ket{c} \\
        &\quad + \left[ \frac{N_1\sqrt{N_2}}{N^{3/2}} e^{i(1+N_2/N_1)t} 2 \cos \left( \frac{\Delta E}{2} t \right) - \frac{\sqrt{N_2}(N_1 - N_2)}{N^{3/2}} \right] \ket{d},
\end{align*}
where
\[ \Delta E = E_2 - E_1 = 2\sqrt{\frac{k_1 N_2}{N_1 N}}. \]
Taking the norm-square of each amplitude, the probability of finding the particle at each type of vertex at time $t$ is
\begin{align}
    p_a(t) &= \frac{4N_1 N_2}{N^2} \sin^2 \left( \frac{\Delta E}{2} t \right), \label{eq:s-prob-time} \\
    p_b(t) &= 0, \nonumber \\
    p_c(t) &= \frac{4N_1N_2^2}{N^3} \cos^2 \left( \frac{\Delta E}{2} t \right) + \frac{4N_1N_2(N_1-N_2)}{N^3} \cos \left( \frac{\Delta E}{2} t \right) \cos \left[ \left( 1 + \frac{N_2}{N_1} \right) t \right] + \frac{N_1(N_1-N_2)^2}{N^3}, \nonumber \\
    p_d(t) &= \frac{4N_1^2N_2}{N^3} \cos^2 \left( \frac{\Delta E}{2} t \right) - \frac{4N_1N_2(N_1-N_2)}{N^3} \cos \left( \frac{\Delta E}{2} t \right) \cos \left[ \left( 1 + \frac{N_2}{N_1} \right) t \right] + \frac{N_2(N_1-N_2)^2}{N^3}. \nonumber
\end{align}
\end{widetext}
As a check, in \fref{fig:s-prob-time-gamma-a}, the dot-dashed green curve corresponds to $\gamma = 0.002$, which is roughly our derived  critical value of $\gamma = 1/N_1 = 1/512 = 0.00195$ We have reproduced this dot-dashed green curve in \fref{fig:s-prob-time-exact-asymptotic}, and we have also added a plot of the asymptotic result $p_a(t)$ from \eqref{eq:s-prob-time} as a solid black curve, and we see strong agreement. As another check, note that $p_a(t) + p_b(t) + p_c(t) + p_d(t) = 1$, since the total probability of the particle being at any vertex must be 1. We see that for large $N_1$ and $N_2$, there is asymptotically no probability of finding the particle at the marked $b$ vertices, but there is probability of finding it at the marked $a$ vertices, and this success probability reaches a maximum at time
\[ t_* = \frac{\pi}{\Delta E} = \frac{\pi}{2} \sqrt{\frac{N_1 N}{k_1 N_2}}, \]
at which the state of the system is
\begin{align*}
    \ket{\psi(t_*)}
        &\approx 2i \frac{\sqrt{N_1N_2}}{N} e^{i(1+N_2/N_1)t_*} \ket{a} \\
        &\quad + \frac{\sqrt{N_1}(N_1 - N_2)}{N^{3/2}} \ket{c} \\
        &\quad - \frac{\sqrt{N_2}(N_1 - N_2)}{N^{3/2}} \ket{d},
\end{align*}
and the probability at each type of vertex is
\begin{align*}
    p_a(t_*) &= \frac{4N_1 N_2}{N^2}, \displaybreak[0] \\
    p_b(t_*) &= 0, \displaybreak[0] \\
    p_c(t_*) &= \frac{N_1(N_1-N_2)^2}{N^3}, \displaybreak[0] \\
    p_d(t_*) &= \frac{N_2(N_1-N_2)^2}{N^3}.
\end{align*}
Thus, the success probability only comes from the $a$ vertices, and it does not reach 1, but rather $p_a(t_*)$ above. This proves that the signless Laplacian exhibits properties of both the Laplacian and adjacency quantum walks, namely only building up success probability at the marked vertices in one partite set and not the other, and having a maximum success probability less than 1, respectively. These results are summarized in the fifth row of \tref{table:summary}.

As a check, in \fref{fig:s-prob-time-exact-asymptotic}, the success probability reaches a maximum at our derived runtime of $t_* = (\pi/2)\sqrt{N_1N/(k_1N_2)} = (\pi/2)\sqrt{512\cdot768/(3\cdot256)} \approx 35.54$, at which the success probability reaches our derived maximum of $p_a(t_*) = 4N_1N_2/N^2 = 4(512)(256)/768^2 \approx 0.889$.


\subsection{Second Degenerate Case}

In case (iii), $\gamma = 1/N_2$, which makes $\ket{b}$ and $\ket{u}$ degenerate eigenvectors of $H^{(0)}$ with eigenvalue $-1 - N_1/N_2$. This case can be analyzed by noting the symmetry of the complete bipartite graph, which allows us to swap the labels of the partite sets, i.e., $a \leftrightarrow b$, $c \leftrightarrow d$, $k_1 \leftrightarrow k_2$, and $N_1 \leftrightarrow N_2$. With these swaps, we get $\gamma = 1/N_1$ with $\ket{a}$ and $\ket{u}$ degenerate eigenvectors of $H^{(0)}$ with eigenvalue $-1 - N_2/N_1$, which is precisely the previous case (ii) that we already analyzed. Then, all the results from the last section carry over, but with the labels swapped. This is summarized in the final row of \tref{table:summary}.

As a check, in \fref{fig:s-prob-time-gamma-b}, the dashed red curve corresponds to $\gamma = 0.004$, which is roughly our derived value of $\gamma = 1/N_2 = 1/256 = 0.00391$, and at roughly our derived runtime of $t_* = (\pi/2)\sqrt{N_2N/(k_2N_1)} = (\pi/2)\sqrt{256\cdot768/(5\cdot512)} \approx 13.77$, the success probability roughly reaches our derived maximum of $p_b(t_*) = 4N_1N_2/N^2 = 4(512)(256)/768^2 \approx 0.889$.


\section{\label{section:better-initial}A Better Initial State and Search Algorithm}

As discussed in the Introduction and summarized in the first two rows of \tref{table:summary}, the Laplacian quantum walk, starting with the uniform superposition $\ket{s}$, searches the complete bipartite graph with a success probability of 1. As discussed in \cite{Wong19}, $\ket{s}$ has two properties that motivate its use:

\begin{enumerate}
    \item   It reflects that every vertex is equally likely to be marked by uniformly guessing each of them.
    
    \item   If if we walk without querying the oracle, we are not learning any new information. $\ket{s}$ reflects this by not evolving, since it is an eigenvector of the Laplacian, which we call $\ket{s_L}$ in the first row of \tref{table:initial}.
\end{enumerate}

In contrast, $\ket{s}$ only retains the first property under the adjacency quantum walk and loses the second, since $\ket{s}$ is not an eigenvector of the adjacency matrix $A$. As summarized in the third row of \tref{table:summary}, an adjacency quantum walk starting in $\ket{s}$ searches the complete bipartite graph with a success probability less than $1$. Instead, the eigenvectors of the adjacency matrix retain the second property, but lose the first property since they are not uniform states. One particular eigenvector, $\ket{s_A}$, is given in the second row of \tref{table:initial}, and it causes the adjacency quantum walk to search with a success probability of 1, as summarized in the fourth row of \tref{table:summary}.

Motivated by this, we note that $\ket{s}$ is similarly not an eigenvector of the signless Laplacian $Q$, and evolving from it does not result in a success probability of 1, as summarized in the fifth and sixth rows of \tref{table:summary}. So, let us explore the eigenvectors of $Q$ in hopes of finding an initial state that causes the signless Laplacian quantum walk to search the complete bipartite graph with probability 1. We will see that we do find such an initial state.

\begin{figure}
\begin{center}
    \includegraphics{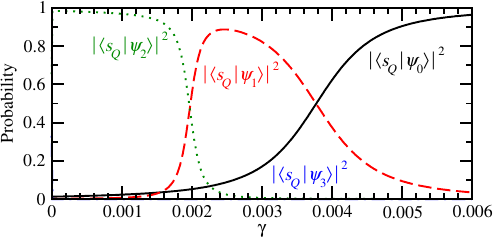}
    \caption{\label{fig:overlaps-sQ}For the complete bipartite graph of $N_1 = 512$ and $N_2 = 256$ vertices with $k_1 = 3$ and $k_2 = 5$ marked vertices, the norm-square of the overlaps of the eigenvectors of the search Hamiltonian with $\ket{s_Q}$. The curve for $|\braket{s_Q}{\psi_3}|^2$ is near zero for all values of $\gamma$.}
\end{center}
\end{figure}

In the $\{ \ket{a}, \ket{b}, \ket{c}, \ket{d} \}$ basis, the signless Laplacian is
\[ Q = \begin{pmatrix}
	N_2 & \sqrt{k_1k_2} & 0 & \sqrt{k_1 N_{k2}} \\
	\sqrt{k_1k_2} & N_1 & \sqrt{k_2 N_{k1}} & 0 \\
	0 & \sqrt{k_2 N_{k1}} & N_2 & \sqrt{N_{k1}N_{k2}} \\
	\sqrt{k_1 N_{k2}} & 0 & \sqrt{N_{k1}N_{k2}} & N_1 \\
\end{pmatrix}, \]
where $N_{ki} = N_i - k_i$ as before, and the eigenvectors and eigenvalues of $Q$ are
\begin{align*}
    & \ket{v_0} = \frac{1}{\sqrt{N}} \left( \sqrt{k_1} \ket{a} - \sqrt{k_2} \ket{b} \right. \\
    &\qquad\qquad\qquad \left. + \sqrt{N_{k1}} \ket{c} - \sqrt{N_{k2}} \ket{d} \right), && 0, \displaybreak[0] \\
    & \ket{v_1} = \frac{1}{\sqrt{N_2}} \left( \sqrt{N_{k2}} \ket{b} - \sqrt{k_2} \ket{d} \right), && N_1, \displaybreak[0] \\
    & \ket{v_2} = \frac{1}{\sqrt{N_1}} \left( \sqrt{N_{k1}} \ket{a} - \sqrt{k_1} \ket{c} \right), && N_2, \displaybreak[0] \\
    & \ket{v_3} = \frac{1}{\sqrt{N}} \left( \sqrt{\frac{k_1 N_2}{N_1}} \ket{a} + \sqrt{\frac{k_2 N_1}{N_2}} \ket{b} \right. \\
    &\qquad\qquad\qquad \left. + \sqrt{\frac{N_{k1}N_2}{N_1}} \ket{c} + \sqrt{\frac{N_{k2}N_1}{N_2}} \ket{d} \right), && N.
\end{align*}
Let us focus on the last eigenvector, $\ket{v_3}$, which we will call $\ket{s_Q}$ going forward because it is the initial state that causes the search to asymptotically succeed with probability 1. To see this graphically, \fref{fig:overlaps-sQ} plots the norm-square of the inner product of $\ket{s_Q}$ with the eigenvectors of the search Hamiltonian, as a function of $\gamma$. Together with \fref{fig:overlaps-a}, we see that when $\gamma = 0.002$, the first and excited states are each roughly half $\ket{s_Q}$ and half $\ket{a}$, and so the system evolves from $\ket{s_Q}$ to $\ket{a}$, as we will analytically prove next. Similarly, from \fref{fig:overlaps-sQ} and \fref{fig:overlaps-b}, we see that when $\gamma = 0.004$, the ground and first excited states are each roughly half $\ket{s_Q}$ and half $\ket{b}$, so the system evolves from $\ket{s_Q}$ to $\ket{b}$, which we will also prove next.

To begin our proof, let us show that for large $N_1$ and $N_2$, $\ket{s_Q}$ is asymptotically $\ket{u}$, which was defined in \eqref{eq:eigensystem-H0}:
\begin{align*} 
    \ket{s_Q}
        &\approx \frac{1}{\sqrt{N}} \Bigg( \sqrt{\frac{k_1 N_2}{N_1}} \ket{a} + \sqrt{\frac{k_2 N_1}{N_2}} \ket{b} \\
        &\qquad\qquad + \sqrt{N_2} \ket{c} + \sqrt{N_1} \ket{d} \Bigg) \\
        &\approx \frac{1}{\sqrt{N}} \left( \sqrt{N_2} \ket{c} + \sqrt{N_1} \ket{d} \right) \\
        &= \ket{u}.
\end{align*}
When $\gamma = 1/N_1$, we proved in \eqref{eq:eigensystem-gamma1} that the first and second excited states of the search Hamiltonian \eqref{eq:H-search-4D} are asymptotically half $\ket{a}$ and half $\ket{u}$, and so
\[ \ket{u} = \frac{1}{\sqrt{2}} \ket{\psi_1} - \frac{1}{\sqrt{2}} \ket{\psi_2}. \]
Then, at time $t$, this state evolves to
\begin{align*}
    \ket{\psi(t)} 
        &= \frac{1}{\sqrt{2}} e^{-iE_1t} \ket{\psi_1} - \frac{1}{\sqrt{2}} e^{-iE_2t} \ket{\psi_2} \displaybreak[0] \\
        &= \frac{1}{\sqrt{2}} e^{-iE_1t} \frac{1}{\sqrt{2}} \left( \ket{a} + \ket{u} \right) \\
        &\quad - \frac{1}{\sqrt{2}} e^{-iE_2t} \frac{1}{\sqrt{2}} \left( \ket{a} - \ket{u} \right) \displaybreak[0] \\
        &= \frac{1}{2} \left( e^{-iE_1t} - e^{-iE_2t} \right) \ket{a} \\
        &\quad + \frac{1}{2} \left( e^{-iE_1t} + e^{-iE_2t} \right) \ket{u} \displaybreak[0] \\
        &= \frac{1}{2} e^{i(1+N_2/N_1)t} \left( e^{i\Delta Et/2} - e^{-i\Delta Et/2} \right) \ket{a} \\
        &\quad + \frac{1}{2} e^{i(1+N_2/N_1)t} \left( e^{i\Delta Et/2} + e^{-i\Delta Et/2} \right) \ket{u} \displaybreak[0] \\
        &= e^{i(1+N_2/N_1)t} i \sin \left( \frac{\Delta E}{2} t \right) \ket{a} \\
        &\quad + e^{i(1+N_2/N_1)t} \cos \left( \frac{\Delta E}{2} t \right) \ket{u} \displaybreak[0] \\
        &= e^{i(1+N_2/N_1)t} i \sin \left( \frac{\Delta E}{2} t \right) \ket{a} \\
        &\quad + e^{i(1+N_2/N_1)t} \cos \left( \frac{\Delta E}{2} t \right) \sqrt{\frac{N_2}{N}} \ket{c} \\
        &\quad + e^{i(1+N_2/N_1)t} \cos \left( \frac{\Delta E}{2} t \right) \sqrt{\frac{N_1}{N}} \ket{d}.
\end{align*}
Taking the norm-square of each amplitude, the probability at each type of vertex is
\begin{align}
    p_a(t) &= \sin^2 \left( \frac{\Delta E}{2} t \right), \label{eq:sQ-prob-time} \\
    p_b(t) &= 0, \nonumber \\
    p_c(t) &= \frac{N_2}{N} \cos^2 \left( \frac{\Delta E}{2} t \right), \nonumber \\
    p_d(t) &= \frac{N_1}{N} \cos^2 \left( \frac{\Delta E}{2} t \right). \nonumber
\end{align}
When
\[ t_* = \frac{\pi}{\Delta E} = \frac{\pi}{2} \sqrt{\frac{N_1 N}{k_1 N_2}}, \]
the probability at each type of vertex reaches
\[ p_a(t_*) = 1, \quad p_b(t_*) = 0, \quad p_c(t_*) = 0, \quad p_d(t_*) = 0. \]
Since $\ket{s_Q}$ is asymptotically $\ket{u}$, and $\ket{u}$ asymptotically evolves to $\ket{a}$ at the runtime, the signless Laplacian quantum walk asymptotically reaches a success probability of 1, and so this is an asymptotically deterministic algorithm. This is summarized in the second-to-last row of \tref{table:summary}.

By swapping $a \leftrightarrow b$, $c \leftrightarrow d$, $k_1 \leftrightarrow k_2$, and $N_1 \leftrightarrow N_2$, when $\gamma = 1/N_2$, the signless Laplacian quantum walk asymptotically evolves to $\ket{b}$ with probability $1$ in time given in the last row of \tref{table:summary}. 

\begin{figure}
\begin{center}
    \subfloat[] {
        \label{fig:sQ-prob-time-gamma-a}
        \includegraphics{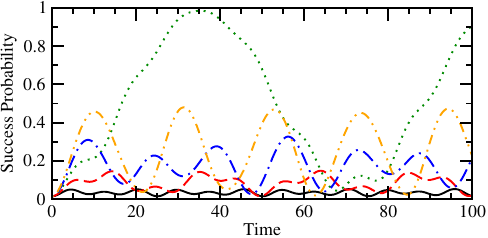}
    }

    \subfloat[] {
        \label{fig:sQ-prob-time-gamma-b}
        \includegraphics{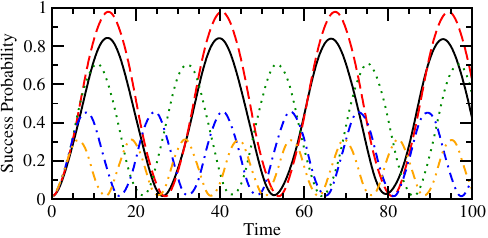}
    }
    \caption{\label{fig:sQ-prob-time-gamma}Success probability versus time for search on a complete bipartite graph of $N_1 = 512$ and $N_2 = 256$ vertices with $k_1 = 3$ and $k_2 = 5$ marked vertices, starting with $\ket{s_Q}$ and using a signless Laplacian quantum walk with various jumping rates. (a) The solid black curve is $\gamma = 0.001$, the dashed red curve is $\gamma = 0.0015$, the dotted green curve is $\gamma = 0.002$, the dot-dashed blue curve is $\gamma = 0.0025$, and the dot-dot-dashed orange curve is $\gamma = 0.003$. (b) The solid black curve is $\gamma = 0.0035$, the dashed red curve is $\gamma = 0.004$, the dotted green curve is $\gamma = 0.0045$, the dot-dashed blue curve is $\gamma = 0.005$, and the dot-dot-dashed orange curve is $\gamma = 0.0055$.}
\end{center}
\end{figure}

These results are confirmed in \fref{fig:sQ-prob-time-gamma}, where the evolution of the success probability is plotted when starting in $\ket{s_Q}$ for various values of $\gamma$. This is analogous to \fref{fig:s-prob-time-gamma}, except with a different initial state. The dot-dashed green curve in \fref{fig:sQ-prob-time-gamma-a} shows that the success probability approaches 1 at time $t_* = (\pi/2)\sqrt{N_1N/(k_1N_2)} = (\pi/2)\sqrt{512\cdot768/(3\cdot256)} \approx 35.54$, and the dashed red curve in \fref{fig:sQ-prob-time-gamma-b} shows that it approaches 1 at time $t_* = (\pi/2)\sqrt{N_2N/(k_2N_1)} = (\pi/2)\sqrt{256\cdot768/(5\cdot512)} \approx 13.77$.

\begin{figure}
\begin{center}
    \includegraphics{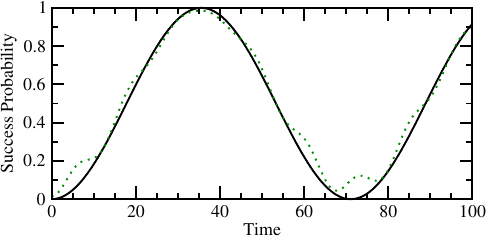}
    \caption{\label{fig:sQ-prob-time-exact-asymptotic}Success probability versus time for search on a complete bipartite graph of $N_1 = 512$ and $N_2 = 256$ vertices with $k_1 = 3$ and $k_2 = 5$ marked vertices, starting with $\ket{s_Q}$ and using a signless Laplacian quantum walk. The dotted green curve is $\gamma = 0.002$ and is copied from \fref{fig:sQ-prob-time-gamma-a}, and the solid black curve is the asymptotic formula $p_a(t)$ derived in \eqref{eq:sQ-prob-time}.}
\end{center}
\end{figure}

Analogously to \fref{fig:s-prob-time-exact-asymptotic}, as a check, in \fref{fig:sQ-prob-time-exact-asymptotic}, we plot the dashed green curve from \fref{fig:sQ-prob-time-gamma-a} along with the asymptotic formula $p_a(t)$ derived in \eqref{eq:sQ-prob-time} and see that there is strong agreement between the two.

Finally, $\ket{s_Q}$ can be rewritten in terms of the individual vertices by substituting in the definitions of $\ket{a}$, $\ket{b}$, $\ket{c}$, and $\ket{d}$, resulting in
\begin{align*}
    \ket{s_Q}
        &= \frac{1}{\sqrt{N}} \left( \sqrt{\frac{N_2}{N_1}} \sum_{\substack{i \in V_1 \\ i \in \text{marked}}} \ket{i} + \sqrt{\frac{N_1}{N_2}} \sum_{\substack{i \in V_2 \\ i \in \text{marked}}} \ket{i} \right) \\
        &\qquad\qquad \left. + \sqrt{\frac{N_2}{N_1}} \sum_{\substack{i \in V_1 \\ i \not\in \text{marked}}} \ket{i} + \sqrt{\frac{N_1}{N_2}} \sum_{\substack{i \in V_2 \\ i \not\in \text{marked}}} \ket{i} \right) \\
        &= \frac{1}{\sqrt{N}} \left( \sqrt{\frac{N_2}{N_1}} \sum_{i \in V_1} \ket{i} + \sqrt{\frac{N_1}{N_2}} \sum_{i \in V_2} \ket{i} \right).
\end{align*}
Thus, all the vertices in the left partite set start off with probability $N_2/(N_1 N)$, and all the vertices in the right partite set start off with probability $N_1/(N_2 N)$, as stated in the last row of \tref{table:initial}.

This results adds to a body of literature suggesting that starting in an eigenvector of the quantum walk operator (without the oracle) may be beneficial to search algorithms. This includes the result of Chakraborty \textit{et al.} \cite{Chakraborty2016}, which explored search on Erd\"os-Renyi random graphs using an adjacency quantum walk, and they showed that by starting in the principal eigenvector of the adjacency matrix, the success probability asymptotically reached $1$. In the setting of discrete-time, coined quantum walks, the abstract search algorithm \cite{AKR2005} begins in an eigenvector of the quantum walk operator, although this success probability may not approach 1. Similarly, a discrete-time quantum walk searching the complete bipartite graph \cite{Wong31} evolves more smoothly from an eigenvector of the quantum walk operator than from the uniform state, but this success probability evolves around $1/2$ rather than $1$.


\section{\label{section:runtimes}Runtime Comparison}

In this section, we compare the runtimes of the various quantum walk search algorithms. From \tref{table:summary}, for each type of quantum walk, the deterministic search algorithm has the same runtime as the nondeterministic search algorithm. Since the nondeterministic search algorithm may need to be repeated before finding a marked vertex, its actual total runtime is expected to be longer. Thus, the deterministic search algorithms are better, and these are the ones we will compare. To distinguish them, let us name the runtimes of the deterministic algorithms as:
\begin{align}
    &t_{L,a} = \frac{\pi}{2} \sqrt{\frac{N}{k_1}}, \nonumber \\
    &t_{L,b} = \frac{\pi}{2} \sqrt{\frac{N}{k_2}}, \nonumber \displaybreak[0] \\
    &t_{A} = \frac{\pi}{\sqrt{2}} \sqrt{\frac{N_1 N_2}{k_2N_1 + k_1 N_2}}, \label{eq:runtimes} \displaybreak[0] \\
    &t_{Q,a} = \frac{\pi}{2} \sqrt{\frac{N_1 N}{k_1 N_2}}, \nonumber \\
    &t_{Q,b} = \frac{\pi}{2} \sqrt{\frac{N_2 N}{k_2 N_1}}, \nonumber
\end{align}
where $t_{L,a}$ and $t_{L,b}$ are the runtimes for the Laplacian quantum walk search algorithm to find a marked vertex in the left or right partite set, respectively, $t_A$ is the runtime for the adjacency quantum walk to find a marked vertex in either partite set, and $t_{Q,a}$ and $t_{Q,b}$ are the runtimes for the signless Laplacian quantum walk search algorithm to find a marked vertex in the left or right partite set, respectively.

\begin{figure}
\begin{center}
    \includegraphics{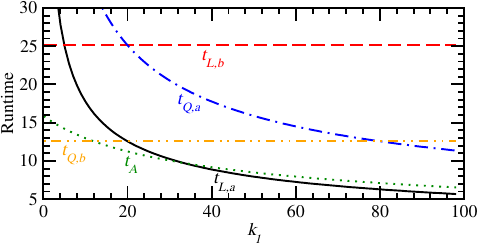}
    \caption{\label{fig:runtimes}Runtimes for search on a complete bipartite graph of $N_1 = 1024$ and $N_2 = 256$ vertices with $k_1$ varying and $k_2 = 5$ marked vertices.}
\end{center}
\end{figure}

\begin{table}
\caption{\label{table:fastest}Parameter regimes for when the Laplacian, adjacency, or signless Laplacian quantum walk searches the complete bipartite graph the fastest.}
\begin{ruledtabular}
\begin{tabular}{cc}
		Conditions & Fastest Quantum Walk \\[0.05in]
        \hline & \\[-0.1in]
		\multicolumn{1}{l}{$N_1 \gg N_2$:} \\[0.05in]
        $k_1 < \frac{k_2N_1(N_1-N_2)}{N_2 N}$ & $t_{Q,b}$ Fastest \\[0.1in]
		$\frac{k_2N_1(N_1-N_2)}{N_2 N} < k_1 < \frac{k_2 N_1 N}{N_2(N_1-N_2)}$ & $t_A$ Fastest \\[0.1in]
		$k_1 > \frac{k_2 N_1 N}{N_2(N_1-N_2)}$ & $t_{L,a}$ Fastest \\[0.1in]
        \hline & \\[-0.1in]
		\multicolumn{1}{l}{$N_1 \ll N_2$:} \\[0.05in]
        $k_2 < \frac{k_1N_2(N_2-N_1)}{N_1 N}$ & $t_{Q,a}$ Fastest \\[0.1in]
		$\frac{k_1N_2(N_2-N_1)}{N_1 N} < k_2 < \frac{k_1 N_2 N}{N_1(N_2-N_1)}$ & $t_A$ Fastest \\[0.1in]
		$k_2 > \frac{k_1 N_2 N}{N_1(N_2-N_1)}$ & $t_{L,b}$ Fastest \\[0.05in]
\end{tabular}
\end{ruledtabular}
\end{table}

Let us compare these runtimes when $N_1 \gg N_2$. First, the runtimes are plotted in \fref{fig:runtimes} with varying values of $k_1$ when searching a complete bipartite graph where $N_1 \gg N_2$. We observe that the dotted green curve ($t_A$) is always faster than the dashed red ($t_{L,b}$) and dot-dashed blue ($t_{Q,a}$) curves. To prove the first of these, note that $t_A$ is maximized when $k_1 \to 0$, at which it takes a value of
\[ t_{A,\text{max}} = \frac{\pi}{\sqrt{2}} \sqrt{\frac{N_2}{k_2}}, \]
and this is always smaller than $t_{L,b}$ since $N_2 < N/2$. To prove the second relationship,
\begin{align*}
    t_{Q,a} 
        &= \frac{\pi}{2} \sqrt{\frac{N_1 (N_1 + N_2)}{k_1 N_2}} > \frac{\pi}{2} \sqrt{\frac{N_1 (N_1 + N_2)}{k_2N_1 + k_1 N_2}} \\
        &> \frac{\pi}{2} \sqrt{\frac{N_1 (N_2 + N_2)}{k_2N_1 + k_1 N_2}} = \frac{\pi}{\sqrt{2}} \sqrt{\frac{N_1 N_2}{k_2N_1 + k_1 N_2}} \\
        &= t_A,
\end{align*}
and so $t_A < t_{Q,a}$. Since $t_A$ is faster than both $t_{L,b}$ and $t_{Q,a}$, we can ignore the latter two in determining which algorithm is the fastest, and just compare $t_A$ with the remaining runtimes $t_{L,a}$ and $t_{Q,b}$. Numerically, in \fref{fig:runtimes}, we observe that the dot-dot-dashed orange curve ($t_{Q,b}$) is fastest for small values of $k_1$, the dot-dashed green curve ($t_A$) is fastest for intermediate values of $k_1$, and the solid black curve ($t_{L,a}$) is fastest for large values of $k_1$. Analytically, we prove this by comparing the respective runtimes from \eqref{eq:runtimes}, yielding that $t_{Q,b}$ is faster than $t_A$ when $k_1 < k_2N_1(N_1-N_2)/(N_2 N)$, and $t_A$ is faster than $t_{L,a}$ when $k_1 < k_2 N_1 N / [N_2(N_1-N_2)]$. As a check, for the graph in \fref{fig:runtimes}, these transitions occur when $k_1 = 5\cdot1024(1024-256)/(256\cdot1280) = 12$ and $k_1 = 5\cdot1024\cdot1280/(256(1024-256)) = 100/3 \approx 33.33$. These results are summarized in the first set of rows in \tref{table:fastest}.

By swapping $a \leftrightarrow b$, $k_1 \leftrightarrow k_2$, and $N_1 \leftrightarrow N_2$, the runtime comparison is obtained when $N_2 \gg N_1$, and the results are summarized in the last set of rows of \tref{table:fastest}.

We see that for some parameter regimes, the signless Laplacian quantum walk is the fastest of the three quantum walk search algorithms.


\section{\label{section:conclusion}Conclusion}

We have conducted the first algorithmic investigation of a quantum walk governed by the signless Laplacian by analyzing how it searches the complete bipartite graph with multiple marked vertices. When the graph is irregular, the signless Laplacian behaves differently from the individual Laplacian and adjacency quantum walks by combining the features of them, and it may search more quickly depending on the parameters. This contrasts with the non-algorithmic investigation of \cite{Alvir2016}, which showed that Laplacian and signless Laplacian quantum walks---without the searching component/oracle---have identical probability distributions when evolving on bipartite graphs. This raises the opportunity that the signless Laplacian could be a new tool in developing faster quantum algorithms based on quantum walks.


\begin{acknowledgments}
	Thanks to Jonathan P.~Wrubel for useful discussions about layered ferromagnetic materials with antiferromagnetic interactions. This material is based upon work supported in part by the National Science Foundation \mbox{EPSCoR} Cooperative Agreement OIA-2044049, Nebraska’s EQUATE collaboration. Any opinions, findings, and conclusions or recommendations expressed in this material are those of the author(s) and do not necessarily reflect the views of the National Science Foundation.
\end{acknowledgments}


\bibliography{refs}

\end{document}